\pdfoutput=1
\documentclass[pagenumbers]{usenixconf}
\usepackage{textweaks}
\usepackage{comment}
\usepackage{upgreek}
\usepackage{url}
\usepackage{xspace}
\usepackage[small,bf]{caption}
\usepackage{listings}
\usepackage{graphicx}
\usepackage{booktabs}
\usepackage[compact]{titlesec}
\usepackage{algorithm}
\usepackage{algpseudocode}
\usepackage{amsmath}
\usepackage[svgnames,table]{xcolor}
\usepackage{tikz}
\usepackage{pgfplots}
\pgfplotsset{
    ylabel right/.style={
        after end axis/.append code={
            \node [rotate=90, anchor=north] at (rel axis cs:1,0.5) {#1};
        }   
    }
}
\usepackage{authblk}

\usepackage{subfigure}

\newcommand{\StatexIndent}[1][3]{%
  \setlength\@tempdima{\algorithmicindent}%
  \Statex\hskip\dimexpr#1\@tempdima\relax}

\newcommand{\us}{$\upmu$s\xspace}
\newcommand{\name}{Memshare\xspace}

\title{Memshare: a Dynamic Multi-tenant Memory Key-value Cache}

\begin{document}

\author[1]{Asaf Cidon}
\author[2]{Daniel Rushton}
\author[3]{Stephen M Rumble}
\author[2]{Ryan Stutsman}
\affil[1]{Stanford University}
\affil[2]{University of Utah}
\affil[3]{Google Inc}

\maketitle{}

\section*{Abstract}

Web application performance is heavily reliant on the hit rate of memory-based caches.
Current DRAM-based web caches statically partition their memory across multiple applications sharing the cache.
This causes under utilization of memory which negatively impacts cache hit rates.
We present \name, a novel web memory cache that dynamically manages memory across applications.
\name provides a resource sharing model that guarantees private memory to different applications
while dynamically allocating the remaining shared memory to optimize overall hit rate.
Today's high cost of DRAM storage and the availability of high performance CPU and memory bandwidth,
make web caches memory capacity bound.
\name's log-structured design allows it to provide significantly higher hit rates
and dynamically partition memory among applications
at the expense of increased CPU and memory bandwidth consumption.
In addition, \name allows applications to use their own eviction policy for their objects,
independent of other applications.
We implemented \name and ran it on
a week-long trace from a commercial memcached provider.
We demonstrate that \name increases the combined hit rate of the
applications in the trace by an 6.1\% (from 84.7\% hit rate to 90.8\% hit rate)
and reduces the total number of misses by
39.7\% without affecting system throughput or latency.
Even for single-tenant applications,
\name increases the average hit rate of the current state-of-the-art memory cache
by an additional 2.7\% on our real-world trace.

\section{Introduction}

DRAM-based caches have become essential for reducing application
latency and absorbing massive database request loads in web applications. Facebook has dozens of applications
that access hundreds of terabytes of data stored in memcached~\cite{memcached} in-memory
caches~\cite{ramcloud-recovery}. Smaller
companies are using outsourced multi-tenant in-memory caches
to cost-effectively boost SQL database
performance.

High access rates and slow backend database performance mean that reducing
the miss rate directly translates to significant end-to-end application
performance. For example, one Facebook memcached pool achieves a 98.2\% hit
rate~\cite{atikoglu2012workload}. With an average cache latency of 200~\us and
MySQL access times of 10~ms, increasing the hit rate to 99.0\%
(a miss reduction of 44\%)
reduces latency by over 20\% (from 376~\us to 298~\us) and reduces database read load
by 1.8$\times$.  The end-to-end speedup is even greater for user queries, which
often wait on hundreds of reads~\cite{nishtala2013scaling}.

Today, operators na\"{\i}vely divide memory statically
across applications.  For
example, Facebook, which manages its own data centers and cache
clusters~\cite{nishtala2013scaling,atikoglu2012workload}, physically partitions
machines into separate cache pools for isolation.  Similarly,
Memcachier~\cite{memcachier,dynacache}, which supplies a cache-as-a-service for hundreds of customers,
statically designates a portion of each cache server's memory
for a specific customer. 

Static partitioning achieves strong isolation, but is inefficient
when applications under utilize their memory
while others are short of resources. Moreover, it is
difficult for cache operators to decide how much memory should be allocated
to each application over time.
Ideally, a web cache should automatically learn and assign the optimal memory partitions
for each application based on their changing working sets;
if an application needs a short term memory boost, it should
be able to borrow memory from one that needs it less.

To this end, we design {\em \name}, a novel
multi-tenant DRAM cache that improves cache hit rates by
exploiting shared and idle memory resources while providing performance isolation guarantees.
To facilitate dynamic partitioning of memory among applications, \name
stores each application's objects in a segmented in-memory log. \name
uses an \emph{arbiter} to dynamically decide which applications require
more memory and which applications are over-provisioned, and it uses a
\emph{cleaner} to evict objects based
on their rank and to compact memory to eliminate fragmentation.
\name enables resource sharing for varying sharing models.

This paper makes two main contributions:
\begin{enumerate}
\itemsep0em
\item \name is the first multi-tenant web memory cache that provides
isolation guarantees, similar to applications' private memory allocation,
while optimally utilizing shared and idle memory.
\name achieves this with a novel sharing model for caches that relies on
dynamic and automatic profiling and adaptive memory reallocation
to boost overall hit rate.
\item \name uniquely enforces isolation through a log-structured design with application-aware cleaning while retaining
fungibility of memory between applications that have objects of different sizes.
Due to its memory efficient design, \name achieves \emph{significantly} higher hit rates
than the state-of-the-art memory cache, both in multi-tenant environments and in single-tenant environments.
\end{enumerate}

In \name, each application specifies a minimum amount of
private memory, and the remaining shared memory is used flexibly to maximize hit
rate. \name maximizes the hit rate by estimating the current hit rate gradient~\cite{cliffhanger}
of each application and awarding memory to the application that would benefit
from it the most. This enables cache providers to
significantly increase hit rates with fewer memory resources while insulating
individual applications from slowdowns due to sharing.  Operators can also run
all applications without private memory shares, in which case \name will
automatically determine an effective allocation for each application that
balances overall hit rate.  Even when all memory is partitioned among
applications, \name can increase overall system efficiency without affecting
performance isolation by allowing idle memory to be reused between applications.

For all sharing models, \name lets applications specify their preferred
eviction policy (e.g., LRU, LFU, Segmented LRU). The eviction policy is expressed
using a \emph{ranking function}~\cite{beckmann2016modeling}. For example, in order to implement LRU,
objects are ranked based on the timestamp of their last access. To implement LFU,
objects are ranked based on their access frequency.

Existing memory caches cannot support these properties; they typically use a
slab allocator~\cite{documentation,dynacache,cliffhanger}, where objects
of different sizes are assigned to slab classes and eviction is done
independently on a class-by-class basis. This greatly limits the ability to reassign
memory from one application to another or reassign memory between objects of different size.

\name replaces the slab allocator with a novel log-structured
allocator that makes memory fungible between objects of different sizes and
applications.  The main drawback of the log-structured allocator is that
in order to reassign memory, it continously repacks memory contents, which increases CPU and memory
bandwidth utilization.  However, trading off increased hit rates with higher CPU
and memory bandwidth utilization is an attractive option, since web memory
caches are typically memory capacity bound and not CPU bound.  For example,
in a week-long trace from Memcachier, cache inserts result in 
less than 0.0001\% memory bandwidth utilization and similarly neglible CPU overhead.
CPU and memory bandwidth should be viewed as under utilized
resources that can be leveraged to increase the cache efficiency,
which motivates the use of a log-structured design for memory caches.

More evidence comes from
Nathan Bronson from the data infrastructure team at Facebook: ``Memcached shares a RAM-heavy server
configuration with other services that have more demanding CPU requirements, so
in practice memcached is never CPU-bound in our data centers. Increasing CPU to
improve the hit rate would be a good trade off.''~\cite{nathan}.
Even under the worst case scenario of high CPU load, \name's
cleaner can dynamically shed load by giving up eviction policy accuracy.  In
practice, it can strongly enforce global eviction policies like LRU with minimal
CPU load.

We implement \name and analyze its performance by running a week-long trace
from Memcachier, a multi-tenant memcached service~\cite{dynacache}.
We show that \name adds 6.1\% to the overall cache hit rate compared to
memcached.
We demonstrate that this miss reduction does not affect overall system throughput for real workloads, since CPU and
memory bandwidth are significantly under utilized. Our
experiments show that \name achieves its superior hit rates and
consuming less than 10~MB/s of memory bandwidth, even under aggressive settings. This represents
only about 0.01\%
of the memory bandwidth of a single CPU socket.
We demonstrate that in the case of a single-tenant application running
in the cache, \name increases the number of hits by extra 2.37\% compared to
Cliffhanger~\cite{cliffhanger}, the state-of-the-art single-tenant cache.
In conclusion, to the best of our knowledge, \name achieves significantly higher
average hit rates both for multi-tenant and single-tenant applications than any other memory cache.

\section{Motivation}

Memory-based web caches have become an essential part of the infrastructure of web applications.
Companies like Facebook, Twitter, Dropbox, and Box maintain clusters of thousands of dedicated servers that run web caches like memcached~\cite{memcached} that serve a wide variety of real-time and batch applications. Smaller companies utilize caching-as-a-service providers such as ElastiCache~\cite{elasticache}, Redis Labs~\cite{redis-labs} and
Memcachier~\cite{memcachier}. These multi-tenant cache providers may split a single server's memory among dozens or hundreds of applications.

Today, cache providers partition the memory space statically across multiple applications
sharing the same cache server. For example, Facebook, which manages its own
cache clusters, partitions its applications to a handful of pools~\cite{nishtala2013scaling,atikoglu2012workload}.
These pools are clusters of memcached servers that cache objects with similar QoS requirements.
The choice of which applications are allocated into each pool is manual.
Caching-as-a-service providers such as Memcachier~\cite{memcachier,dynacache} allow
their customers to purchase a certain amount of memory.
Each application is statically allocated memory on several
memcached servers, and these servers maintain a separate eviction queue for
each application.

%These three applications exhibit widely different behaviors.
%Application 1 has a very high rate of compulsory misses (over 25\% of the requests), which
%represent objects that are only accessed once. Application 2 has a very predictable set of requests
%that achieves a high hit rate with LRU, and application 3 has a bursty diurnal access pattern. Section~\ref{sec:e2e} compares results on larger mixes of applications.

\subsection{Partitioned vs Shared}

We compare two different resource sharing schemes with memcached:
the static partitioning used by Memcachier, and a greedy shared memory policy, both using memcached's slab allocator with LRU.
In the static partitioning, we run applications just as they run in our commercial Memcachier trace;
each is given isolated access to the same amount of memory it had in the
trace. In the shared policy, applications share all memory, and their objects share eviction queues.
An incoming object from any application evicts objects from the tail of the shared
per-class eviction queues, regardless of which application's objects are
disposed. 
We use a motivating example of three different applications (3, 5 and 7) selected from a
week-long trace of memcached traffic running on Memcachier.
These applications suffer from bursts of requests,
so they clearly demonstrate the trade offs between the partitioned and shared memory policies. 

\definecolor{asparagus}{rgb}{0.53, 0.66, 0.42}
\definecolor{applegreen}{rgb}{0.55, 0.71, 0.0}
\definecolor{alizarin}{rgb}{0.82, 0.1, 0.26}

\begin{table}[t]
\centering
\footnotesize
\begin{tabular}{rrrrr}
\toprule
& \multicolumn{2}{c}{Hit Rate} \\
App & Partitioned & Shared \\ \midrule
3 &  97.6\% & \cellcolor{alizarin!25} 96.6\%  \\
5 &  98.8\% & \cellcolor{applegreen!25} 99.1\% \\
7 &  30.1\% & \cellcolor{applegreen!25} 39.2\% \\
\midrule
Combined &  87.8\% &  88.8\% \\

\bottomrule
\end{tabular}
\caption{Average hit rate of Memcachier's partitioned and shared policy over a week.}
\label{tbl:static}
\end{table}

% Other important facts:
% $ tail -n 1 *util*memcached* | grep -v == | grep -v '^$' | awk '{print $2}' | sort -rn | head -n 8                                                               
%
% Memcachier median app memory utilization: 70.4126%
% memached median app memory utilization: 91.7462%

\begin{figure}[t]
\centering
\includegraphics[width=\columnwidth,natwidth=288,natheight=216]{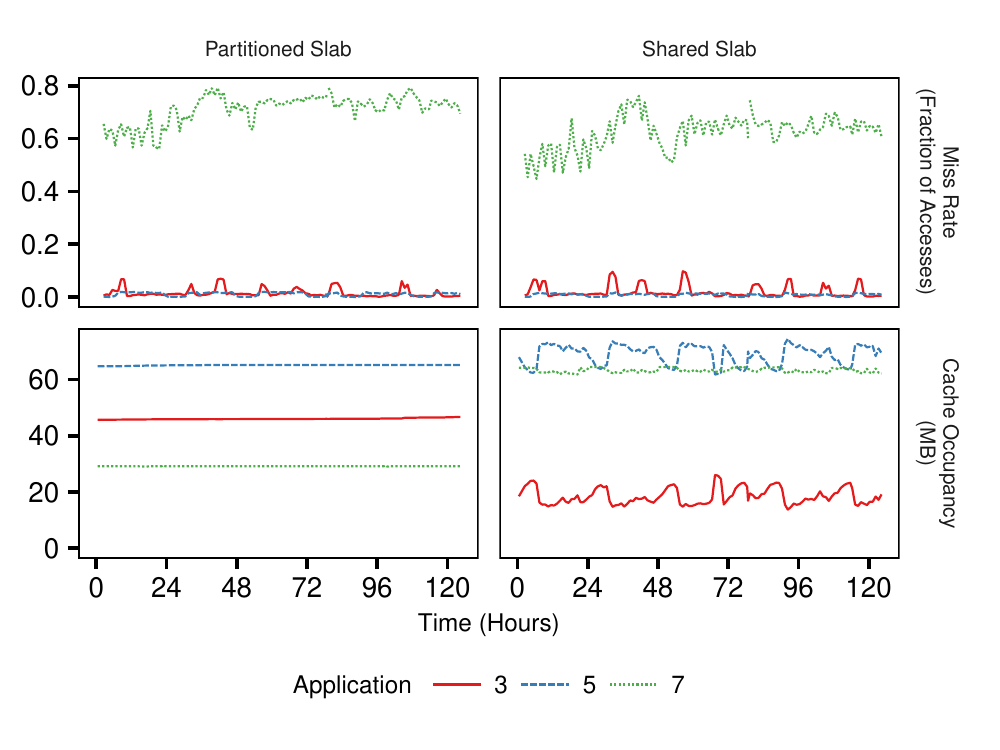}
\caption{Miss rate and cache occupancy of Memcachier's partitioned and shared policies over time.}
\label{fig:static}
\end{figure}

Table~\ref{tbl:static} shows the average hit rates over a week of the three applications in
both configurations. Figure~\ref{fig:static} depicts the average miss rate and cache occupancy over an entire week.
The shared policy gives a superior
overall hit rate; however, it comes at
the expense of a 1\% drop in application~3's hit rate.
This would result in 43\% higher database load and increased latencies for that application.
Notice that the figure shows that the greedy scheme significantly changes the memory allocation
between the applications; application~3 loses about half its memory, while
application~7 doubles its share.

% The greedy
% policy strongly deviates from the amount of memory that was allocated by the static policy.
% We capture the notion of memory allocation fairness using Jain's Fairness Index (JFI)~\cite{JFI}:

% \begin{equation*}
% \small
% \begin{aligned}
%   & \dfrac{\Big(\sum\limits_{i=1}^n \dfrac{\mbox{\em targetMem}(i)}{\mbox{\em actualMem}(i)}\Big)^2}{n \cdot \sum\limits_{i=1}^n \Big(\dfrac{\mbox{\em targetMem}(i)}{\mbox{\em actualMem}(i)}\Big)^2}
% \end{aligned}
% \end{equation*}

% Where $n$ is the number of applications we are running in the cache, $targetMem$ is the average
% amount of memory that each application is supposed to get and $actualMem$ is the average amount of memory
% each application was actually allocated. JFI is 1 when memory allocation is perfectly fair; when allocation is completely unfair JFI is $\frac{1}{n}$. Fairness in terms of memory allocation is important, since memory allocation
% is a common way to measure resource allocation in web memory caches. For example, in Memcachier,
% customers buy a certain amount of guaranteed reserved memory from the cache provider.

\subsection{Slab Allocation Limits Multi-tenancy}

Ideally, a multi-tenant eviction policy should combine the best of partitioned and shared
resource sharing. It should provide performance isolation and it should also allow applications to claim unused memory
resources when appropriate, so that an application that has a burst of requests would be able to temporarily
acquire such resources. This raises two key requirements for the policy.
First, it must be able to dynamically arbiter which applications can best
benefit from additional memory and which applications will suffer the least
when losing memory.
Second, it needs to be
able to dynamically re-allocate memory across applications.

Unfortunately, memory allocators like memcached's slab allocator would greatly limit the ability
to move memory between applications, since objects of different sizes are partitioned
in their own slabs. The following example illustrates the problem.

Assume that a cache decided 
to move 4~KB of memory from application 1 to application 3.
In the trace, application 1's median object size is 56 bytes, and the median object size for application 3 is
576 bytes. In Memcachier, each 1~MB slab of memory is assigned a {\em size class}; the slab is divided into fixed
sized chunks according to its class. Classes are in units of $64 \times 2^i$ up to 1~MB (i.e. 64~B, 128~B, $\ldots$, 1~MB).
Each object is stored in the smallest class that can contain the object.
Therefore, objects of 56~B are stored in a 1~MB
slab of 64~B chunks, and 576~B are stored in a
1~MB slab of 1~KB chunks.

There are two problems with moving memory across applications in a slab allocator.
First, since memory is split into 1~MB slabs, if we want to move 4~KB of memory from application 1 to 3,
and application 3's objects are on average much larger than application 1's objects,
we have to move an entire 
1~MB slab of memory from application 1. Therefore,
application 1 would have to evict 1~MB full of small objects,
some of which may be hot as well as cold items; memcached tracks LRU rank via an explicit list, which doesn't relate to how objects are physically grouped within slabs.
Second, the newly allocated 1~MB memory for application 3 could only be used for a single object size.
So for example, application 3 would only be able to use it for objects of size 256-512 byte
or objects between 512-1024 bytes. If it needed to be allocated more memory for objects
of both sizes, it would need application 1 to evict yet another slab.
Ideally, the cache should be able to evict only the bottom ranked items from application 1, based
on application 1's eviction policy, which have a total size of 4~KB.

This motivates a new design for a multi-tenant cache memory allocator, which
can dynamically move variable amounts of memory among applications, while
preserving applications' eviction policy and priorities.
\section{Design}
\label{sec:design}

This section presents the design of \name
\name is a lookaside cache server that supports
the memcached API.
When the cache is full, its goal is to evict stale objects to make
room for fresh values. \name dynamically and automatically assigns
a portion of the cache to each application, while monitoring how
effectively each application uses its share, and it reapportions memory to
improve hit rates.

\name is split into two key components. First, \name's arbiter must
determine how much memory should be assigned to each application (its {\em
targetMem}). Second, \name's cleaner implements these assignments
by prioritizing eviction from
applications that are using too much cache space.

\begin{figure}[t]
\begin{minipage}{\columnwidth}
  \includegraphics[width=\columnwidth,natwidth=752.212524,natheight=194.400009]{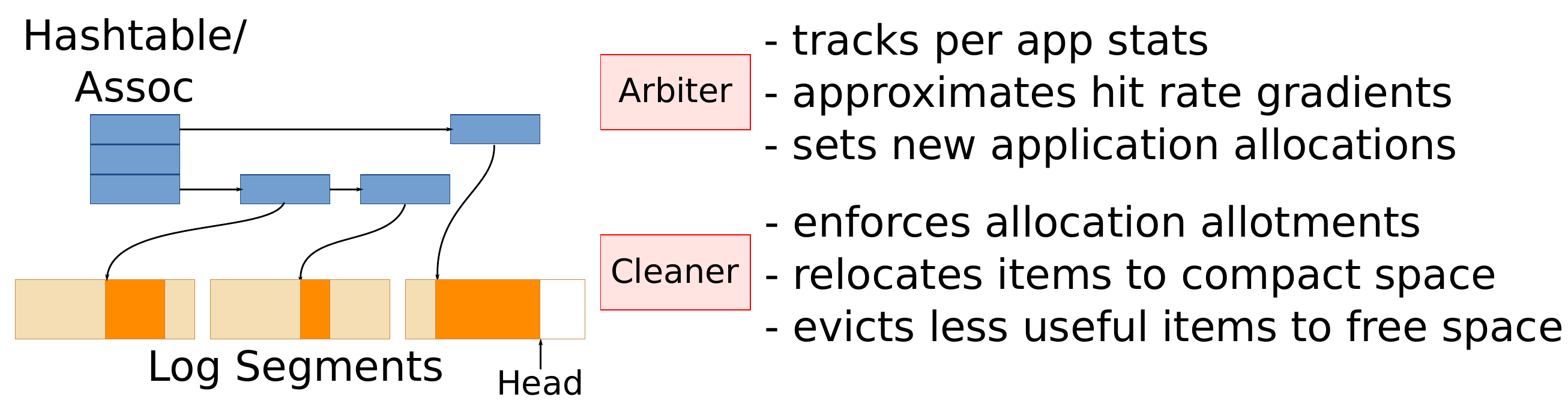}
\end{minipage}
\caption{The \name design. Incoming items are allocated from the head of a segmented in-memory
log. The hash table maps keys to their location in the log. The arbiter
monitors operations and sets allocation policy. The cleaner evicts
items according to the arbiter's policy and compacts free space.}
\label{fig:design}
\end{figure}

\subsection{The Cleaner and Arbiter}

\name's in-memory cleaner fluidly reallocates memory among applications.
The cleaner finds and evicts the least useful items for any
application from anywhere in memory, and it coalesces the resulting free space
so that it can be used to host any object from any application of any size.
This coalescing also provides fast allocation and high memory utilization.
% Conventional memory allocators like memcached's slab allocator can reallocate
% space, but they hurt the cache hit rate.  Memory is assigned to size classes
% (or application classes in this case), so whenever more memory is needed by one
% class it must be taken from another. This reassignment happens in large slabs
% (1~MB) for efficiency, and all of the items in a reassigned slab have to first
% be evicted. Unfortunately, the items in a slab are mixed with respect to
% LRU-rank and usefulness to the application. This means reallocating a slab
% necessarily evicts both cold {\em and} hot items. \name's cleaner avoids this
% problem.

All items in \name are stored in a segmented in-memory log
(Figure~\ref{fig:design}). New items are allocated contiguously from the same
active {\em head} segment, which starts empty and fills front-to-back.  Once an
item has been appended to the log, the hash table entry for its key is pointed
to its new location in the log.  Unlike slab allocator systems like memcached,
\name's segments stores objects of all sizes from
all applications; they are all freely intermixed.  By default, segments are
1~MB; when the head segment is full, an empty ``free'' segment is chosen as
head.

When the system is running low on free segments ($< 1\%$ of total DRAM), it
begins to run the cleaner in the background, in parallel with handling normal
requests. The cleaner frees space in two steps.  First, it evicts objects that belong to an application
that is using too much cache memory.  Second, it compacts free space together
into whole free segments by moving objects in memory.  Keeping a small pool of free segments allows the
system to tolerate bursts of writes without blocking on cleaning.

\name relies on its arbiter to choose which objects the cleaner should prefer
for eviction. To this end we define the {\em need} of each application as its need for memory:

\begin{equation*}
\begin{aligned}
  \mbox{\em need}(\mbox{\em app}) = \dfrac{\mbox{\em targetMem}(\mbox{\em app})}{\mbox{\em actualMem}(\mbox{\em app})}
\end{aligned}
\end{equation*}

Where {\em actualMem} is the actual number of bytes currently storing objects
belonging to the application, and {\em targetMem} is the number of bytes
that the application is supposed to be allocated. In the case of
partitioned resource allocation {\em targetMem} is constant.
If the need of an application is above 1, it means it needs to be
allocated more memory. Similarly, if the need is below 1, it is
consuming more memory than it is supposed to have.

The arbiter ranks applications according to their {\em need} for
memory, and the cleaner prefers to clean from segments that contain more data
from the applications that have the lowest need. Items in a segment being cleaned are
considered one-by-one; some are saved and others are evicted.

\begin{algorithm} [t!]
\caption{Memory relocation}
\label{alg:evict}
{\small
\begin{algorithmic}[1]
\Function{cleanMemory}{segments, n}
	\State{relocated = 0}
	\State{residual = (n - 1) $\cdot$ segmentSize}
	\While{relocated $<$ residual}
		\State{app = arbiter.maxNeed()}
		\State{object = maxRank(segments, app)}
		\State{segments.remove(object)}
		\If{object.size $\leq$ residual - relocated}
			\State{relocate(object)}
			\State{relocated = relocated + object.size}
			\State{app.actualMem = app.actualMem + object.size}
		\Else
			\State{break}
		\EndIf
	\EndWhile
\EndFunction
\end{algorithmic}
}
\end{algorithm}
\noindent

Cleaning is composed of ``passes''. Each pass takes $n$ distinct segments and 
outputs at most $n - 1$ new distinct segments, freeing up at least one empty segment.
This is done by writing back the most essential objects into the $n - 1$ output segments.
The writing is done contiguously so that free space, caused by obsolete objects that were overwritten,
is also eliminated.
$n$ is a system parameter that is discussed in Section~\ref{sec:evaluation}.
Note that multiple passes can run in parallel.

In each pass, \name selects a fraction of the segments for cleaning randomly and a fraction
based on which segments have the most data from applications with
the lowest need.  This directs the cleaner to choose more segments occupied by
applications that are using more than their fair share. Random selection helps
to avoid pathologies. For example, if segments were only chosen based on
application need, some applications might be able to remain over provisioned
indefinitely so long as there are worse offenders. We chose the fraction to be equal to 0.5
following experimentation with the Memcachier traces.

\begin{figure}[t!]
\begin{minipage}{\columnwidth}
\includegraphics[width=\columnwidth,natwidth=676.799988,natheight=194.400009]{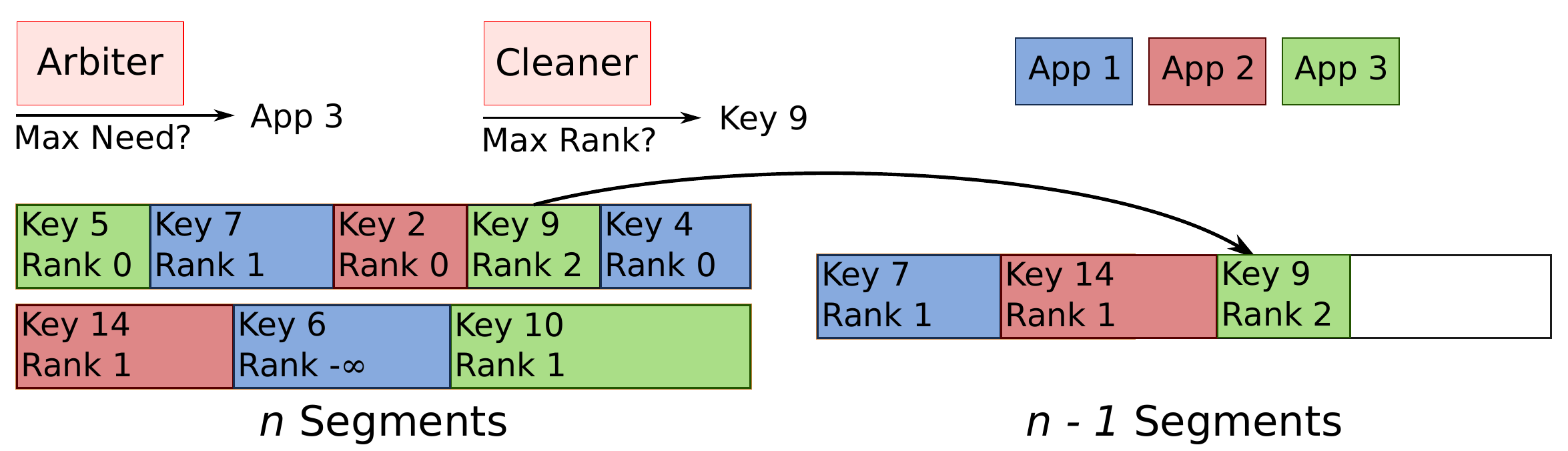}
\end{minipage}
\caption{\name relocates objects from $n$ segments to $n-1$ segments. The the arbiter first chooses the application with the highest need, and the cleaner relocates the object with the highest rank among the objects of that application.}
\label{fig:eviction}
\end{figure}

Once a set of segments is selected for cleaning, the cleaner ranks all of the
items in the segments to determine which items should be preserved and which
items should be evicted.  Figure~\ref{fig:eviction} and
Algorithm~\ref{alg:evict} show how this is done in a single cleaning pass.
{\em segments} is a list of all the items from the segments that are candidates
for eviction. In order to choose which item to relocate
next, the cleaner first determines the application that has the highest need
({\em maxNeed}). Among the items in the segments that belong to that
application, the cleaner then chooses the item with the highest rank ({\em
maxRank}, e.g. LRU-rank). It relocates the item by copying it and updating its entry in the hash table.
After the item is relocated,
the need for that
application is recalculated. The
process is repeated until the $n - 1$ segments are full or all items are relocated. The remaining
items are evicted by dropping them from the hash table, and the need for the applications' whose objects were evicted is adjusted accordingly.

\name can use any generic ranking function on items
to prioritize them for eviction; in fact, it can be determined by the
application.  \name supports any ranking function $\mbox{\em rank}(t, f)$, that
is based on the timestamp $t$ of the last access of each item and $f$ the
number of times it has been accessed. For example, in order to implement LRU,
the ranking function is $\mbox{\em rank}(t) = t$, or simply equal to the time
stamp of the last access of the item. LFU is implemented as $\mbox{\em rank}(f)
= f$ as the number of times an item has been accessed. Segmented LRU can be
implemented as a combination of the time stamp of the last access of the item
and the number of times it has been accessed. Throughout the paper, when
evaluating the hit rate of different caches, we use LRU as the default eviction
policy.

A key idea behind \name's design is that memory partitioning is enforced
by the decision of which items to clean, while any application can write
at any time to the cache. Consider the case where \name is configured
by the cache operator to enforce a static partitioning among applications, and one application continuously
writes new items to the cache, while other applications do not write new items.
Since the memory allocation is partitioned, {\em targetMem} will remain constant. As the first application
writes new items to the cache,
its {\em actualMem} will increase until its need drops below the need
of all other applications.
When the memory fills and cleaning starts, the arbiter 
will choose to clean data from the application
that has the lowest need and will begin to evict its data.
If there are other active applications competing for memory, this application's {\em actualMem} will drop, and its need will
increase appropriately.
%The {\em targetMem} of an application is a notion of its fair share,
%and the goal of the arbiter is to be as close as possible
%to the fair share of an application.

\subsection{Balancing Eviction Accuracy and Cleaning}

The performance of \name involves a trade off between the accuracy of the eviction
policy, determined by the parameter $n$ and the rate of updates to the cache.
$n$ is limited by the rate of updates
to the cache. The higher the rate of updates, the faster the cleaner must be to
free up enough memory to keep up with the updates. Section~\ref{sec:e2e}
evaluates this cost and finds for our trace the cleaning cost is less than
0.01\% utilization for single modern CPU socket.
Even so,
the cleaner can be made faster and cheaper by decreasing $n$; decreasing $n$ reduces the amount of the data the cleaner will rewrite to reclaim a segment worth of free space.
This also results in the eviction of items that are ranked higher by their respective
applications, so the accuracy
of the eviction policy decreases. In our design, $n$ can be dynamically adjusted
based on the rate of updates to the memory cache. Note that memory cache workloads
typically have a low update rate (less than 3\%)~\cite{nishtala2013scaling}.

The last segment out of the $n-1$ segments produced by the cleaning pass may be under utilized,
because of many dead items in the original $n$ segments. One of the interesting properties of the cleaner,
is that the $n-1$ segments are sorted based on need and rank. Therefore, a further optimization of the cleaner
is to delete the last segment in case its utilization is low (e.g., under 50\%), since it contains
the lowest rank and need objects of the $n-1$ segments.

\section{\name's Sharing Model}

\name provides a resource allocation policy that provides both
a minimum amount of private memory for each application, and assigns the rest of the
cache's shared memory to the application that would benefit from it the most in terms of hit
rate. Each application is allocated a
certain amount of private guaranteed memory ({\em privateMem}). Any remaining
memory on the server is shared among the different applications, and we
refer to it as {\em sharedMem}. At each point in time, \name has a target amount of memory
it is trying to allocate to each application, {\em targetMem}.
In the case of statically partitioned memory, {\em sharedMem} is zero,
and {\em targetMem} is always equal to {\em privateMem} for each application.

\definecolor{asparagus}{rgb}{0.53, 0.66, 0.42}
\definecolor{applegreen}{rgb}{0.55, 0.71, 0.0}
\definecolor{alizarin}{rgb}{0.82, 0.1, 0.26}

\begin{table}[t]
\centering
\footnotesize
\begin{tabular}{rrrrr}
\toprule
& \multicolumn{2}{c}{Hit Rate} \\
App & Partitioned & \name 50\% \\ \midrule
3 &  97.6\% & \cellcolor{applegreen!25} 99.4\% \\
5 &  98.8\% & \cellcolor{applegreen!25} 98.8\% \\
7 &  30.1\% & \cellcolor{applegreen!25} 34.5\% \\
\midrule
Combined & 87.8\% & 89.2\% \\
\bottomrule
\end{tabular}
\caption{Average hit rate of \name with 50\% private memory compared to the partitioned policy.}
\label{tbl2}
\end{table}

% Other important facts:
% $ tail -n 1 *util*memcached* | grep -v == | grep -v '^$' | awk '{print $2}' | sort -rn | head -n 8                                                               
%
% Memcachier median app memory utilization: 70.4126%
% memached median app memory utilization: 91.7462%

{\em targetMem} defines its application's fair share.
Therefore, the resource allocation policy needs to ensure that
each application's {\em targetMem} does not drop below its {\em privateMem},
and that the remaining {\em sharedMem} is distributed among each application
in a way that maximizes some performance goal such as the maximum overall hit rate.

In maximizing the overall hit rate among the difference applications, we can estimate each application's hit rate curve,
which is the hit rate it would achieve for a given amount of memory. If we know
one of the application's hit rate curves, we can allocate memory
to the applications that would achieve the highest number of hits~\cite{dynacache}. However, estimating
the entire hit rate curve for each application running on the cache can be expensive
and inaccurate~\cite{cliffhanger}.

Instead of estimating the entire hit rate curves, we estimate the local hit rate curve gradient,
by leveraging \emph{shadow queues}. 
A shadow queue is an extension of the cache that does not store
the values of the items, only the keys. Each application has its own
shadow queue. Objects are evicted from the
cache into the shadow queue. For example, assume a certain application has 10,000 objects
stored in the cache, and it has a shadow queue that stores the keys of 1,000 objects.
When a request misses the eviction queue but hits the application's shadow queue, it
means that if the application was allocated space for another 1,000 objects,
the request would have been a hit.
The rate of hits in the shadow queue
provides a local approximation of an application's hit rate curve gradient~\cite{cliffhanger}.
Therefore, the application with the highest rate of hits in its shadow queue
would provide the highest number of hits if its memory was incrementally increased.

\begin{algorithm} [t!]
\caption{Shared memory: set target memory}
\label{alg:assign}
{\small
\begin{algorithmic}[1]
\Function{setTarget}{request, application}
	\If{request $\not\in$ cache AND \par
		\hskip\algorithmicindent request $\in$ application.shadowQueue}
		\State{candidateApps = \{\}}
		\For{app $\in$ appList}
			\If{app.sharedMem $\geq$ credit}
				\State{candidateApps = candidateApps + \{app\}}
			\EndIf
		\EndFor
		\State{pick = pickRandom(candidateApps)}
		\State{application.sharedMem =  \par
		\hskip\algorithmicindent application.sharedMem + credit}
		\State{pick.sharedMem = pick.sharedMem - credit}
	\EndIf
	\For{app $\in$ appList}
		\State{app.targetMem = \par
		\hskip\algorithmicindent app.privateMem + app.sharedMem}
	\EndFor
\EndFunction
\end{algorithmic}
}
\end{algorithm}
\noindent

Algorithm~\ref{alg:assign} depicts how we decide to set the {\em targetMem}.
Each application is initially assigned a proportion of {\em sharedMem}.
When a request enters the cache, if it is a miss,
we check to see if its key is stored in its application's shadow queue,
i.e., if the request hit the application's shadow queue. If there was
a shadow queue hit, that application is assigned a {\em credit}. A credit
represents a certain amount of memory (e.g., 64~KB) from the total pool of shared memory.
The algorithm takes
away a credit at random from an application.
{\em pickRandom} is a function that randomly chooses an object
from a list. We will show below how the cleaner
utilizes {\em targetMem} to choose which applications to evict objects from.
{\em appList} is a list of all applications in the cache and {\em cache}
is a list of all objects in the cache.

Table~\ref{tbl2} compares \name with the statically partitioned scheme used by Memcachier.
We ran \name when 50\% of the memory that was allocated in the original Memcachier trace of each application
is used as private memory and the rest is allocated
as shared memory. \name provide near-equal or greater hit rate than the partitioned memcached policy. Even
though 50\% of its memory is reserved, \name also
achieves a higher overall hit rate (89.2\%) than the greedy shared memory scheme (88.8\%).

\begin{table}[t!]
\centering
\footnotesize
\begin{tabular}{rrr}
\toprule
Private Memory & Total Hit Rate \\ \midrule
0\% & 89.4\% \\
 25\% & 89.4\% \\
 50\% & 89.2\% \\
 75\% & 89.0\% \\
 100\% & 88.8\% \\

\bottomrule
\end{tabular}
\caption{Comparison of \name's total hit rate with different amounts of private memory for applications 3, 5, and 7.}
\label{tbl:minmem}
\end{table}

% Other important facts:
% $ tail -n 1 *util*memcached* | grep -v == | grep -v '^$' | awk '{print $2}' | sort -rn | head -n 8                                                               
%
% Memcachier median app memory utilization: 70.4126%
% memached median app memory utilization: 91.7462%

\begin{figure*}[t!]
\centering
\includegraphics[width=0.9\textwidth,natwidth=576,natheight=198]{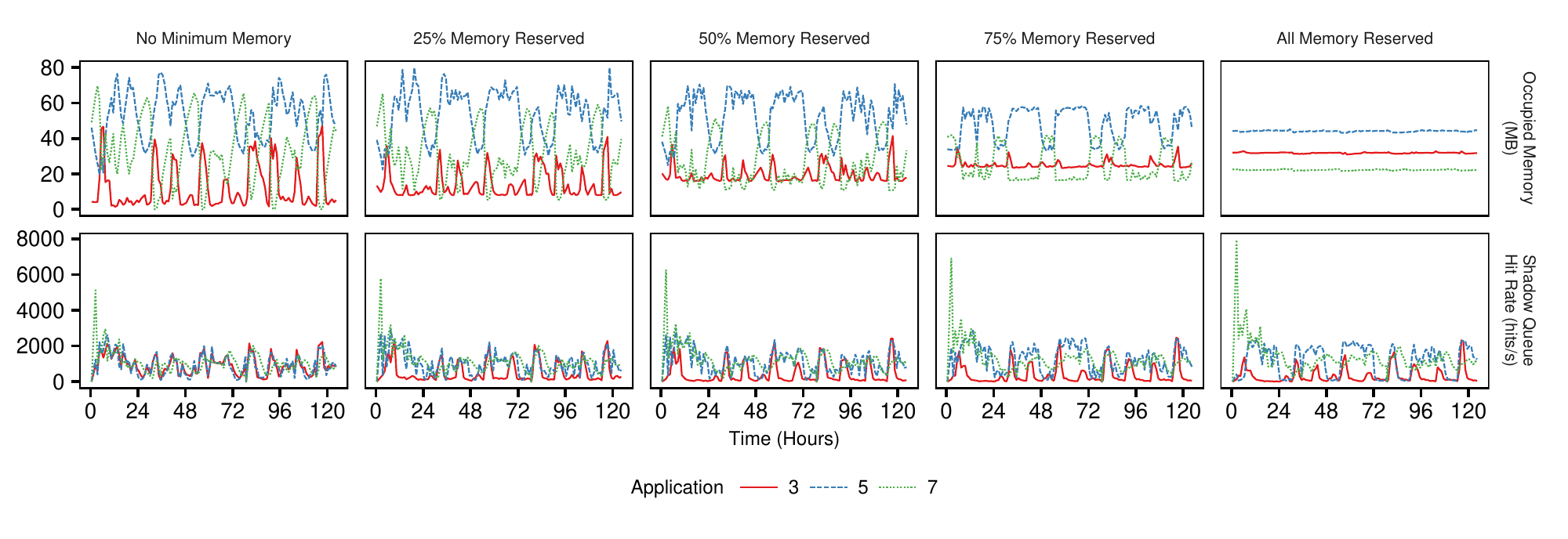}
\caption{Comparison of \name's memory consumption and the rate of shadow queue hits with different amounts of memory reserved for applications 3, 5 and 7. \name assigns more shared memory to applications that have a high rate of shadow queue hits.}
\label{fig:minmem}
\end{figure*}

Table~\ref{tbl:minmem} and Figure~\ref{fig:minmem} further explore the trade off between overall hit rate and
the individual applications hit rates 
as we vary the percentage of memory that is privately reserved. The figure shows how
as the amount of private memory increases, the amount of memory allocated to each application flattens.
In addition, the figure demonstrates that \name's cleaner enforces the private memory
allocation for each application:
the amount of memory allocated to each application does not fall below its private memory reservation.
The figure also shows how \name's memory allocation reacts to the rate of shadow queue hits.
In the far left graphs, when the cache has no reserved private memory, \name allocates
shared memory to the applications that have a high shadow queue hit rate. As \name
allocates more memory to the bursty application, its shadow queue hit rate tempers.
In the far right graphs, when the cache is fully reserved, \name cannot allocate
any additional memory to the bursty applications, and therefore the shadow queue hit rate
is the highest.

\subsection{Prioritizing Shared Memory Allocation Among Applications}

Cache providers may want to guarantee that when certain applications have bursts of requests,
they would get a higher priority than other applications.
In order to accommodate this requirement, \name enables cache operators to assign different shadow queue
credit sizes to different applications. This guarantees that if a certain application
has a higher credit size than other applications, when it requires a larger amount of memory
due to a burst of activity, it will be able to expand its memory footprint faster than
other applications.

\begin{table}[t]
\centering
\footnotesize
\begin{tabular}{rrrrr}
\toprule
  App & Credit Size & Hit Rate & Credit Size & Hit Rate \\ \midrule
3 & 64~KB & 99.4\% & 64~KB & 99.5\%\\
5 & 128~KB & 98.5\% & 64~KB & 98.6\%\\
7 & 192~KB & 33.4\% & 64~KB & 32.3\%\\
\bottomrule
\end{tabular}
\caption{Assigning different credit sizes to each application allows cache operators to prioritize among applications.}
\label{tbl:weights}
\end{table}

% Other important facts:
% $ tail -n 1 *util*memcached* | grep -v == | grep -v '^$' | awk '{print $2}' | sort -rn | head -n 8                                                               
%
% Memcachier median app memory utilization: 70.4126%
% memached median app memory utilization: 91.7462%

Table~\ref{tbl:weights} demonstrates how assigning different weights to different
applications affects their overall hit rate. In this example, application 7 achieves
a higher relative hit rate, since it receives larger credits
in the case of a shadow queue hit.

\subsection{Increasing System Efficiency for Private Memory}

Shared memory is ideal for environments such as Facebook or other large
web-scale providers, where multiple cooperative applications
utilize a shared caching layer, and the operator wants to provide
the best possible overall performance, while providing minimum
guarantees to each application.

However, in some environments, applications are inherently selfish and would like to maximize their private memory,
but the cache operator
still has an incentive to optimize for the highest possible utilization of memory.
If certain applications are under utilizing their private memory, their resources can be re-assigned
without negative impact to their performance.

We present a solution for re-assigning idle memory among applications
that leverages the simple idea of an idle memory tax~\cite{ESX},
where memory that has not been accessed for a long time can be re-assigned.

\begin{algorithm} [t!]
\caption{Idle tax: set target memory}
\label{alg:outsourced}
{\small
\begin{algorithmic}[1]
\Function{setTarget}{app, taxRate, idleTime}
	\State{idleMem = 0}
	\For{object $\in$ app}
		\If{object.timestamp $<$ currentTime - idleTime}
			\State{idleMem $+=$ object.size}
		\EndIf
	\EndFor
	\State{activeFraction = 1 - $\dfrac{idleMem}{app.actualMem}$}
	\State{$\tau$ = $\dfrac{1-activeFraction \cdot taxRate}{1-taxRate}$}
	\State{app.targetMem = $\dfrac{app.privateMem}{\tau}$}
\EndFunction
\end{algorithmic}
}
\end{algorithm}
\noindent

The only difference between \name's shared memory allocation and the idle tax
policy, is in how the arbiter sets each application's {\em targetMem}.
Algorithm~\ref{alg:outsourced} describes how the arbiter computes the
{\em targetMem} for each application in this scenario.
In the algorithm, {\em taxRate} is a number between 0 and 1, which defines what portion of each application's memory
can be reclaimed by other applications if it is idle.
If {\em taxRate} is set to 1, all of the application's
memory can be reclaimed by other applications if the memory is idle (and its {\em targetMem}
will be equal to 0). If {\em taxRate} is equal to 0, the idle tax cache policy is effectively
the same as partitioned allocation.
The algorithm defines idle memory as memory that has not been accessed by a time
interval greater than {\em idleTime}.
The arbiter keeps track of what fraction of each application's memory is idle,
and it sets the {\em targetMem} based on what fraction of each application is idle and
on the tax rate.

In this algorithm, {\em targetMem} can never be higher than {\em privateMem}. Therefore,
if multiple applications do not have any idle memory, and they are competing
for additional memory, the resources will allocated to them on a fair share 
based on their {\em privateMem}. For example, if two applications are contending for 10~MB
of free memory, and one of them has a {\em targetMem} of 5~MB, and the other one
has a {\em targetMem} of 1~MB, the remaining 10~MB will be split in a 5:1 ratio
(8~MB and 2~MB).

\begin{table}[t]
\centering
\footnotesize
\begin{tabular}{rrr}
\toprule
& \multicolumn{2}{c}{Hit Rate} \\
App & Memcachier Partitioned & Idle Tax \\ \midrule
3 & 97.6\% &  99.4\% \\
5 & 98.8\% & 98.6\% \\
7 & 30.1\% & 31.3\% \\
\midrule
Combined & 87.8\% &  88.8\% \\
\bottomrule
\end{tabular}
\caption{Average hit rate of \name's idle tax policy.}
\label{tbl:tax}
\end{table}

% Other important facts:
% $ tail -n 1 *util*memcached* | grep -v == | grep -v '^$' | awk '{print $2}' | sort -rn | head -n 8                                                               
%
% Memcachier median app memory utilization: 70.4126%
% memached median app memory utilization: 91.7462%

Table~\ref{tbl:tax} depicts the hit rate \name's idle tax algorithm
using a tax rate of 50\% and a 5 hour idle time.
In the three application example, the overall hit rate is increased,
because the idle tax cache policy favors objects that have been accessed recently.
Even though application 5's memory suffers a slight decrease,
because some of its idle objects were accessed after more than 5 hours,
this decrease is much more tempered than in the case of greedy shared memory.

\subsection{Automatically Defining Private Memory}

\name's shared memory algorithm tries to optimally distribute
shared memory across applications, while the idle tax algorithm
taxes applications that are not actively using their private memory.
These two approaches can be combined,
for example, in the case where cache operators want to reserve a certain amount of space for an application,
but are not sure how much memory to allocate to each application.
In this case, they can run
\name with the shared memory algorithm using private memory set to 0. After
the algorithm runs for several hours, the cache operator can switch to the idle tax algorithm
and set each application's private memory to be equal to the average target memory of
the shared cache algorithm.

\section{Implementation}
\label{sec:implementation}

%\begin{figure}
%\footnotesize
%\begin{tabular}{lp{4.4cm}}
%\toprule
%\multicolumn{2}{l}{\textbf{Log} -- allocates memory} \\
%\texttt{alloc(size) $\rightarrow$ address} & Return portion of head segment. Request more empty segments from the cleaning if running low.\\
%\texttt{free(address)} & Subtract allocation size from the in-use count for the segment that contained the allocation.\\
%%\midrule
%
%\multicolumn{2}{l}{\textbf{Arbiter} -- monitors app use and sets policy} \\
%\texttt{adjustAppUse} & Foo. \\
%\texttt{mergeCleaningStats} & Foo. \\
%%\midrule
%
%\multicolumn{2}{l}{\textbf{Cleaner} -- reclaims memory, enforces policy via evictions} \\
%\texttt{clean} & Foo. \\
%
%%\midrule
%\multicolumn{2}{l}{\textbf{CleaningPass} -- tracks per} \\
%\texttt{addItem} & Foo. \\
%\texttt{relocateAndEvict} & Foo. \\
%
%\bottomrule
%\end{tabular}
%\caption{The major \name modules and APIs.}
%\label{fig:api}
%\end{figure}

In this section, we describe the implementation of \name and how
its various parts are synchronized.
\name consists of three major modules written in C++
on top of memcached~1.4.24: the log, the arbiter and the cleaner. 
\name reuses most of memcached's units without change including
its hash table, basic transport, dispatch, and request processing.

\subsection{The Log}

The log replaces memcached's slab allocator. It provides a basic \texttt{alloc}
and \texttt{free} interface. On allocation, it returns a pointer to the
requested number of bytes from the current ``head'' segment. If the request is
too big to fit in the head segment, the log selects an empty segment as the new
head and allocates from it.
%When the pool of empty segments runs low the log
%informs the cleaner, which produces additional empty segments
%in the background.
%Freeing an allocation simply decrements the bytes-in-use
%count for the segment that contained the allocation being freed. The
%cleaner uses this information when selecting segments for cleaning. The
%interface is nearly identical to the that of the original slab allocator, so
%integrating the log into memcached requires very few changes.

Allocation of space for new objects
and the change of a head segment are protected by a spinlock. Contention is not a concern since
both operations are inexpensive: allocation increments an offset
in the head segment and changing a head segment requires
popping a new segment from a free list. If there were no
free segments threads would block waiting for the cleaner to add new segments
to the free list. In practice the free list is
never empty (we describe the reason below).

\subsection{The Arbiter}

%The arbiter sets the application sharing policy and implements the ranking
%functions needed by the cleaner to implement the chosen policy.
The arbiter tracks
two key attributes for each application: the amount of space each application is
occupying in the cache and its shadow LRU queue of recently evicted items.  The
SET request handler forwards each successful SET to the arbiter so the
per-application bytes-in-use count can be increased.  On evictions during
cleaning passes, the arbiter decreases the per-application bytes-in-use count
and inserts the evicted items' into the application's shadow queue.  In
practice, the shadow queue only stores the 64-bit hash of each key and the
length of each item that it contains, which makes it small and efficient. Hash
collisions are almost non-existent and do no harm; they simply result in slight
over-counting of shadow queue hits.

\subsection{The Cleaner}

The cleaner
always tries to keep some amount of free memory available. By default, when the
log notices less than 1\% of memory is free it notifies the cleaner, which
starts cleaning. It stops when at least 1\% is free again. If the cleaner falls
behind the rate at which service threads perform inserts, then it starts new
threads and cleans in parallel. The cleaner can clean more aggressively,
by reducing the number of segments for cleaning ({\em n}), or freeing up more segments
in each cleaning pass. This trades
eviction policy accuracy for reduced CPU load and memory bandwidth.
%Alternatively, cleaner threads can instead
%evict more aggressively instead of relocating objects; this trades eviction
%policy accuracy for reduced cleaning CPU load.  Such a configuration might make
%sense on servers intended to share resources with compute bound workloads.

% Where all is there sync?
% - Lock ht buckets for get/set, relocate, and liveness check?
% - Must sync free list and seg list on pass start/stop.
% - Update per-app shadow queue at end of cleaning pass.
%   - How? Can do app-by-app grab lock and stuff it all in.

Cleaning passes must synchronize with each other and with normal request
processing. A spin lock protects the list of full segments and the list of empty
segments. They are both manipulated briefly at the start and end of each
cleaning pass to choose segments to clean and to acquire or release free
segments. In addition, the cleaner uses Memcached's fine-grained bucket locks to synchronize hash
table access. The cleaner accesses the hash table to determine item liveness,
to evict items, and to update item locations when they are relocated.

The arbiter's per-app bytes-in-use counts and shadow queues are protected by a spin lock,
since they must be
changed in response to evictions. Each
cleaner pass aggregates the total number of bytes evicted from each application
and it installs the change with a single lock acquisition to avoid the overhead
of acquiring and releasing locks for every evicted item. The shadow queue is
more challenging, since every evicted key needs to be installed in some
application's shadow queue. Normally, any GET that results in a miss should
check the application's shadow queue.  So, blocking operations for the whole
cleaning pass or even just for the whole duration needed to populate it with
evicted keys would be prohibitive. Instead, the shadow queue is protected with
a spin lock while it is being filled, but GET misses use a \texttt{tryLock}
operation. If the \texttt{tryLock} fails, the shadow queue access is ignored.

The last point of synchronization is between GET operations and the
cleaner. The cleaner never modifies the objects that it moves.
Therefore, GET operations do not acquire the lock to the segment list and continue to
access records during the cleaning pass. In rare schedules, this could result in a GET operation
finding a reference in the hash table to a place in a segment that is cleaned before it is actually accessed.
\name uses an epoch mechanism to make this safe.  Each
request/response cycle is tagged at its start with an epoch copied from a global
epoch number. After a cleaning pass has removed all of the references from the
hash table, it tags the segments with the
global epoch number and then increments it.  A segment is
only reused when all requests in the system are from epochs later than
the one it is tagged with.

\subsection{Modularity}

\name maintains a strong separation between the cleaner and the arbiter, even
though eviction order is highly dynamic and requires tight communication
between the modules. To do this, after a cleaning pass chooses segments, it
notifies the arbiter which items are being cleaned. The arbiter ranks them and
then calls back to the cleaner for each item that it would like to keep. If the
relocation is successful the arbiter updates the item's location in the hash
table.  Once the empty segments have been filled with relocated items, the
arbiter removes the remaining entries from the hash table and updates
per-application statistics and shadow queues. In this way, the cleaner is
oblivious to applications, ranking, eviction policy, and the hash table. Its
only task is efficient and parallel item relocation.

\section{Evaluation}
\label{sec:evaluation}

In this section we present the evaluation of \name on the Memcachier traces and a set of microbenchmarks.
To measure the end-to-end performance of \name,
we ran the week-long Memcachier trace on \name.
Since the Memcachier traces have a low load of requests, we also benchmarked our implementation
using the YCSB~\cite{ycsb} workload generator.

Our experiments run on 4-core 3.4~GHz Intel Xeon E3-1230 v5 (with 8~total
hardware threads) and 32~GB of DDR4 DRAM at~2133~MHz.  All experiments
are compiled and run using the stock kernel, compiler, and libraries on
Debian~8.4 AMD64.

\begin{figure*}[t]
\center
\includegraphics[width=0.8\textwidth,natwidth=468,natheight=144]{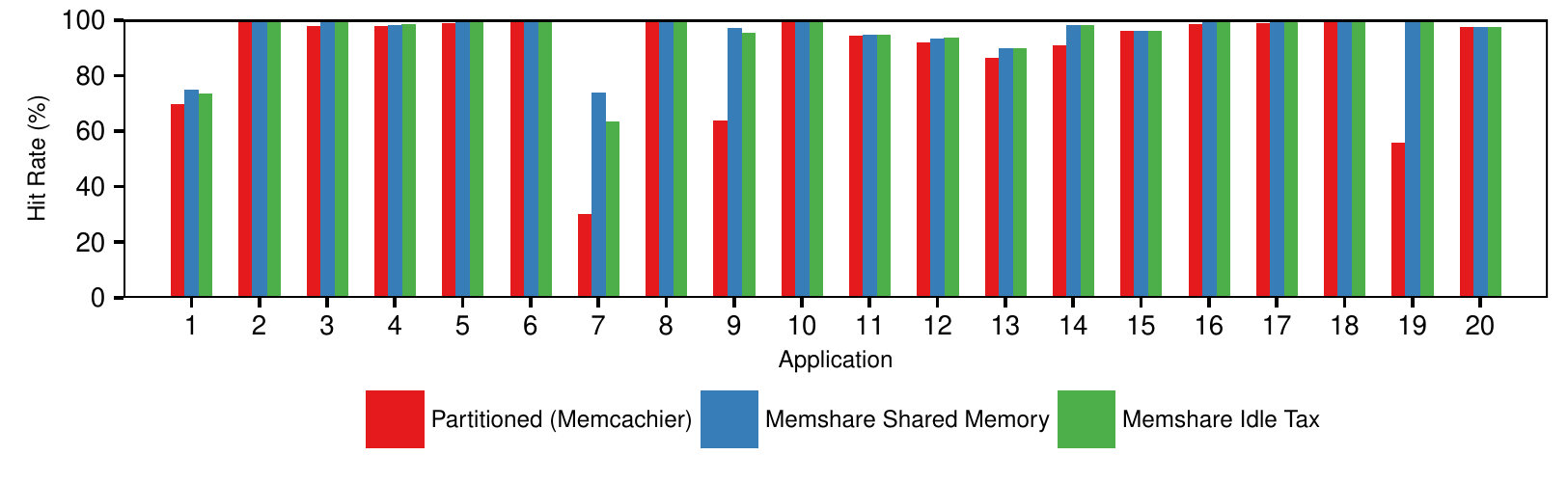}
\caption{Hit rates of \name's shared memory and idle tax algorithms running the top applications of Memcachier, compared to memcached.}
\label{fig:deltas}
\end{figure*}

% Other important facts:
% $ tail -n 1 *util*memcached* | grep -v == | grep -v '^$' | awk '{print $2}' | sort -rn | head -n 8                                                               
%
% Memcachier median app memory utilization: 70.4126%
% memached median app memory utilization: 91.7462%

\begin{table}[t]
\centering
\footnotesize
\begin{tabular}{rrr}
\toprule
Policy & Combined Hit Rate & Miss Reduction\\
\midrule
memcached & 84.66\% & 0.00\% \\
Cliffhanger & 87.73\% & 20.00\% \\
\name Tax & 89.92\% & 34.28\% \\
\name Shared & 90.75\% & 39.69\% \\
\bottomrule
\end{tabular}
\caption{Combined hit rate of \name's idle tax (50\% tax) and shared memory policy (75\% private) compared with Cliffhanger, which is the state-of-the-art slab-based cache and Memcached. The miss reduction column compares the miss rate of the different policies to memcached.}
\label{tbl:hitrate_summary}
\end{table}

\subsection{End-to-end Performance}
\label{sec:e2e}

Our evaluation uses 1~MB segments and 100 candidate segments for cleaning.
We ran the shared cache policy with
75\% of each application's original Memcachier memory as private, and the rest as shared.
For each application we used a shadow queue that represents 10~MB of objects.
We ran the idle tax policy with a 50\% tax rate and with all of the memory
allocated to each application as private.

We tested the end-to-end performance of \name using all the major applications from the Memcachier trace
with the shared memory and idle tax policies.
Figure~\ref{fig:deltas} presents the hit rate results and Table~\ref{tbl:hitrate_summary} presents
the summary.
The shared cache policy provides a higher overall combined hit rate increase,
since it tries to maximize for overall hit rates.

For the shared memory policy, we experimented with using different credit sizes. When \name uses credit sizes that are too small,
shared memory won't be moved quickly enough between applications to maximize hit rate. When it uses credit sizes that are too high,
the allocation of shared memory among applications will oscillate, which will cause excessive evictions. We
found that for the Memcachier applications a credit size of 64~KB provides a good balance.

\begin{table}[t]
\centering
\footnotesize
\begin{tabular}{rrrrr}
\toprule
Segments (n) & Hit Rate & Memory Bandwidth (MB/s) \\
\midrule
1 & 89.20\% & 0.04 \\
10 & 90.47\% & 2.14 \\
20 & 90.58\% & 2.86 \\
40 & 90.74\% & 4.61 \\
60 & 90.74\% & 6.17 \\
80 & 90.75\% & 7.65 \\
100 & 90.75\% & 9.17 \\
\bottomrule
\end{tabular}
\caption{Combined hit rate and memory bandwidth consumption of top 20 applications in Memcachier trace using \name with the shared memory policy with 75\% private memory, and varying the number of segments in each cleaning pass.}
\label{tbl:candidate_segments}
\end{table}

Table~\ref{tbl:candidate_segments} presents the combined hit rate and cleaner memory bandwidth consumption of \name's shared memory policy when
varying $n$, the number of segments that participate in each cleaning pass. The table shows that for the Memcachier traces,
there is a diminishing increase in hit rate beyond n=40.
While memory bandwidth use increases as the number of candidate segments is higher,
near peak hit rates can be achieved for this trace while consuming less than 0.01\% of the memory bandwidth of a single modern CPU socket.
Even at 100 candidate segments, the memory bandwidth
of \name is less than 10 MB/s for the top 20 applications in the trace.

\subsubsection{Single Tenant Hit Rate}
\label{sec:single-app-hit-rate}

\begin{table}[t]
\centering
\footnotesize
\begin{tabular}{rrr}
\toprule
Policy & Average Single Tenant Hit Rate \\
\midrule
memcached & 88.3\% \\
Cliffhanger & 93.1\% \\
\name 100\% Private & 95.5\% \\
\bottomrule
\end{tabular}
\caption{Average hit rate of the top 20 applications in the trace with \name with 100\% private memory, compared with Cliffhanger and memcached. We run each application as a single tenant.}
\label{tbl:avg_hitr}
\end{table}

In addition to providing multi-tenant guarantees, \name's log structured design
significantly improves hit rates on average for individual applications on a cache
which uses a slab allocator.
Table~\ref{tbl:avg_hitr} compares the average hit rates between \name and two systems that utilize slab allocators:
memcached and Cliffhanger~\cite{cliffhanger}. Within a single tenant application, Cliffhanger optimizes the amount of memory allocated to each slab
to optimize for its overall hit rate. However, \name's log structured design provides superior hit rates
compared to Cliffhanger, because it allows memory to be allocated fluidly for objects of different sizes.
In contrast, each time Cliffhanger moves memory from one slab class to another,
it must evict an entire 1~MB of objects, including objects that may be hot.
On average, \name with 100\% private memory
increases the hit rate by 7.13\% compared to memcached and by 2.37\% compared to Cliffhanger.

% Figure~\ref{fig:app19} provides an example of why \name outperforms memcached's
% slab allocator. The figure depicts the hit rate curve of application 19.
% In memcached, objects of different sizes are assigned to slab classes,
% each of which has its own eviction queue. 
% The size of the LRU queue of each slab class is determined greedily
% based on which objects arrive to the server first.
% This means that if the object sizes change over time, such as in the case of application 19, the new objects
% may have LRU queues that are too small. This affects results in cliffs
% in the hit rate curve.
% In contrast, \name approximates a global LRU eviction policy across objects of all sizes,
% and therefore it does not have any cliffs in its hit rate curve.

\subsection{Microbenchmarks}

Since the Memcachier traces do not result in a high CPU utilization, we also
ran microbenchmarks of \name using the YCSB framework~\cite{ycsb}, which
incurs significantly higher CPU and memory bandwidth utilization. The
results show that \name incurs minimal CPU utilization and throughput
overheads. In addition, \name increases read latency and reduces write latency
under load.

For all of our microbenchmarks we used 25~B objects, 23~B keys over 100
million operations. We only measured the overheads when the cleaner was fully
active.

\subsubsection{Latency}

\begin{table}[t]
\centering
\footnotesize
\begin{tabular}{rrrr}
\toprule
& \multicolumn{3}{c}{Latency} \\
& GET Hit & GET Miss & SET \\
\midrule
memcached & 21.44~\us & 21.8~\us & 29.48~\us \\
\name & 22.04~\us & 23.0~\us & 23.62~\us \\
\bottomrule
\end{tabular}
\caption{Average latencies of \name compared to memcached under an artificial workload that causes high CPU utilization. \name's shadow queue lookup increases the latency in the case of GET cache misses.}
\label{tbl:latency}
\end{table}

Table~\ref{tbl:latency} presents the average latency of \name compared to memcached when the
cache is full and the cleaner is running. These numbers are taken with both the
clients and cache server threads running on the same machine. Consequently, they
represent a worst case; typical cache access times are dominated by the network software stack and round trip times~\cite{ramcloud}.
\name's GET hit latency
is 2.7\% higher than memcached. \name incurs a 5.5\% latency overhead for GET misses,
since it checks whether the key exists in the shadow queue. This extra overhead
could be eliminated.
Note that adding an overhead to a GET miss is typically insignificant, since the application
needs to issue a database query, which takes tens to hundreds of milliseconds.

\subsubsection{CPU and Throughput}
\label{sec:cpu-and-tput}

\begin{table}[t]
\centering
\footnotesize
\begin{tabular}{rr}
\toprule
& Ops per Second \\
\midrule
memcached 5\% writes & 705,968 \\
\name 5\% writes & 690,332 \\
memcached 100\% writes & 540,325 \\
\name 100\% writes & 519,277 \\
\bottomrule
\end{tabular}
\caption{Average throughput of \name compared to memcached under a YCSB workload with 5\% writes and 95\% reads and under a worst case workload with 100\% writes.}
\label{tbl:throughput}
\end{table}

Table~\ref{tbl:throughput} compares the throughput of \name with memcached under a YCSB workload with 5\% writes and 95\% reads and under a workload with 100\% writes. On average, \name has a 2.2\% lower throughput for the first workload and a 3.9\% lower throughput under the punishing all writes workload.

Most of the throughput loss is due to \name's cleaner. To quantify the throughput loss,
we measured the CPU time spent by \name on different tasks.
In the 5\% write workload, \name spends 5.1\% of the process's CPU time on cleaning,
and 1.1\% of the process's CPU time testing shadow queues on GET misses.
The 100\% write workload is unrealistic (such a workload does not need a cache), but it highlights the worst case throughput cost for the hit rate improvements that \name gives.
With a 100\% write workload
12.8\% of the process's CPU time is spent on cleaning.

Overall, the small decrease in throughput of \name is well justified. In-memory
caches are typically capacity-bound not throughput-bound, and they often
operate under low loads~\cite{nathan,dynacache}. In particular, the Memcachier
trace introduces loads which are two orders of magnitude lower than the throughput
of \name.

\subsubsection{Memory Overhead and Utilization}

\name has a small memory overhead. The shared memory policy uses shadow queues that store keys
which represent 10~MB of objects. The memory overhead of the shadow queues depends on the size
of the objects. For example, assuming objects are small on average (128~B),
a shadow queue stores 81,920 keys. Only 8~B key hashes are kept, so key length isn't a factor. In this
case, the overhead is 81,920~$\cdot$~8~B~=~640~KB per application.
The rest of the data structures used by \name have a negligible memory overhead.

As mentioned earlier, \name's cleaning process does waste some space by keeping some
segments pre-cleaned. However, this free space only
represents about 1\% of the total cache space in our implementation.  Even with
idling a small fraction of memory, \name is still better than memcached's
slab allocator, since it eliminates the internal fragmentation that slab
allocators suffer from.  For example, in the trace,
memcached's fragmentation causes the cache to run at 70\%-90\% memory utilization.

\section{Related Work}

Our work is inspired by ideas from previous work on memory resource
allocation and caching. Cliffhanger~\cite{cliffhanger} introduced a technique to estimate the local
gradient of hit rate curves in memory caches using shadow queues, for re-balancing slabs that belong
to objects of different sizes. We applied a similar idea to assign memory among
multiple applications. Compared to Cliffhanger, \name achieves significantly higher hit rates and can
flexibly move memory across applications,
because it uses a log-structured memory allocator
rather than a slab allocator.

We were inspired by the idea of taxing idle memory from Carl Waldspurger's work on ESX~\cite{ESX} and
min-funding revocation~\cite{lottery}. In contrast to ESX,
\name keeps track of the timestamps of the last access of all objects, so it is
fairly simple to keep track of which objects are idle.

Our concept of a ranking function to rank the priorities of the objects of each
application was inspired by the concept of ranking functions introduced by
Beckmann et al~\cite{beckmann2016modeling}, as a flexible model for replacement policies
for CPU caches.

RAMCloud~\cite{lsm} and MICA~\cite{mica} have applied the ideas of log-structured
file systems~\cite{rosenblum1992design,seltzer1993implementation,seltzer1995file,blackwell1995heuristic,matthews1997improving} to DRAM-based caches. 
In addition, there are other examples of using log-structured caches in other contexts, such as a CDN photo cache~\cite{RIPQ}
and mobile device caches~\cite{Aghayev}. \name uses a log-structured
design similar to RAMCloud and MICA, but differs from them in several important ways. First,
unlike these systems, \name addresses multi-tenant resource sharing. Second, both RAMCloud and MICA
rely on a FIFO based approach for eviction, which typically suffers from lower hit rates than LRU.
\name enables application developers to
apply any eviction policy using their own ranking functions.

\subsection{Resource Allocation and Sharing}

FairRide~\cite{fairride} provides a general framework for cache memory allocation and
fairness, in particular when applications, processes or threads share data.
While shared data among competing applications is common in certain scenarios,
it is not common in key-value web caches in a data center setting.
For example, in both Facebook and Memcachier, different
applications have their own unique key spaces, and they cannot access the same keys on memcached.
For applications that do not share data, FairRide implements a memory partitioning policy
in a distributed setup.
\name, unlike FairRide, can efficiently utilize non-reserved and allocated idle memory
to optimize the hit rate of applications and provide them with a memory boost
in case of a burst of requests.

Mimir~\cite{mimir} and Dynacache~\cite{dynacache} provide a framework for approximating
the stack distance curves of web memory caches,
in order to understand how much memory needs to be allocated to different applications.
These systems are essentially offline optimizers, since an optimization solver that runs
on historical data does not adapt when application workloads change on the fly.
In addition, they do not provide a mechanism for
allocating memory among different applications sharing the same cache. Mimir's techniques can be
combined with \name to help cache providers make offline decisions about
cluster sizing.

Most previous efforts on cloud resource allocation, such as Moirai~\cite{moirai},
Pisces~\cite{pisces}, DRF~\cite{DRF} and Choosy~\cite{choosy} are focused on performance isolation and sharing
in terms of requests per second (throughput), not in terms of cache hit rate which is the key
ingredient in determining speedup in data center memory caches~\cite{nathan}.

There have been several projects analyzing cache fairness and sharing in the context of multicore
processors~\cite{iyer,kim2014,ubik}. In the context of multicore, fairness is viewed as a function
of total system performance. Unlike CPU caches, DRAM-based web caches are
typically separate from the compute
and storage layer, so the end-to-end performance impact is unknown to the cache.

Ginseng~\cite{ginseng} and RaaS~\cite{raas1,raas2} provide a framework for memory
pricing and auctioning for outsourced clouds, but they only focus on
pricing memory in the case where each application has their own dedicated memory cache server running on a VM.
In contrast, \name enables multiple applications to share the same
memory cache server, without the need to rely on VM isolation.
This is the preferred deployment model for most web application providers (e.g.,
Facebook, Dropbox, Box).

\subsection{Eviction Policies}

Many eviction schemes can be used in conjunction with \name. For example, Greedy-Dual-Size-Frequency~\cite{cherkasova1998improving} takes into account request sizes to replace LRU as a cache eviction algorithm for web proxy caches. Greedy-Dual-Wheel~\cite{li2015gd} outperforms LRU by leveraging the knowledge of how long each request takes to be computed by the database. C-EVA~\cite{beckmann2015bridging} computes the opportunity cost per byte for each object stored in a cache. Other eviction policies, like ARC~\cite{megiddo2003arc}, LRU-K~\cite{o1993lru}, 2Q~\cite{2Q}, LIRS~\cite{Jiang:2002:LEL:511399.511340} and LRFU~\cite{lee2001lrfu,lee1999existence}, offer a combination of LRU and LFU.

\subsection{Memory Cache Performance}

MemC3~\cite{fan2013memc3} and work from Intel Labs~\cite{li2015architecting} improve the throughput of Memcached on multicore, by increasing concurrency and removing lock bottlenecks. While these systems significantly improve the throughput of Memcached, they do not improve overall hit rates. In the case of Facebook and Memcachier, Memcached is memory capacity bound and not CPU bound~\cite{nathan,dynacache}.
\section{Conclusions}

Web cache memory hit rate is one of the most important factors in
determining end-to-end web application performance.
Current web memory caches statically partition memory across applications. This
leaves room for significant improvement in increasing 
the hit rate of applications.
We describe \name, a multi-tenant web memory cache that provides
higher hit rates while
maintaining private memory for each application.
\name's log-structured
design provides a significant hit rate benefit over current
caches that use slab allocation, both in the case of a multi-tenant and single-tenant memory cache.
In addition, \name lets cache operators tune
priorities and private memory across applications, and it allows
applications to implement their own eviction policies.

\footnotesize \bibliographystyle{abbrv}
\bibliography{bib}

\end{document}